\documentclass[sigconf, nonacm]{acmart}

\usepackage{svg}
\usepackage{graphicx, caption}
\usepackage{placeins} %
\usepackage{multicol}
\usepackage{stfloats}
\usepackage{seqsplit}
\usepackage{siunitx}
\usepackage[labelformat=simple]{subcaption}
\begin{document}

\title{DumpKV: Learning based lifetime aware garbage collection for key value separation in LSM-tree}

\author{Zhutao Zhuang}
\affiliation{%
  \institution{Microsoft}
}
\affiliation{%
  \institution{ Sun Yat-sen University}
}
\email{zhuangzhutao@gmail.com}
\email{zhuangzht@mail2.sysu.edu.cn}

\author{Xinqi Zeng}
\affiliation{
\institution{Chongqing University}
}
\email{xinqizeng2002@outlook.com}

\author{Zhiguang Chen}
\affiliation{
\institution{Sun Yat-sen University}
}
\email{chenzhg29@mail.sysu.edu.cn}
\email{
zhiguang.chen@nscc-gz.cn
}

\begin{abstract}
 Key-value separation is used in LSM-tree to stored large value in separate log files to reduce write amplification, but requires garbage collection to garbage collect invalid values. 
 Existing garbage collection techniques in LSM-tree typically adopt static parameter based garbage collection to garbage collect obsolete values which  struggles to achieve low write amplification and it's challenging to find proper parameter for garbage collection triggering. 
 In this work we introduce DumpKV, which introduces learning based lifetime aware garbage collection with dynamic lifetime adjustment to do efficient garbage collection to achieve lower write amplification.
 DumpKV manages large values using trained lightweight model with features suitable for various application based on past write access information of keys to give lifetime prediction for each individual key to enable efficient garbage collection. 
 To reduce interference to write throughput DumpKV conducts feature collection during L0-L1 compaction leveraging the fact that LSM-tree is small under KV separation .
 Experimental results show that DumpKV achieves lower write amplification by 38\%-73\% compared to existing key-value separation garbage collection LSM-tree stores with small feature storage overhead.

\end{abstract}

\keywords{Storage engine, Key-value storage, LSM-tree storage, Machine learning system}

\maketitle

\section{Introduction}

LSM(Log-structure merge)-tree \cite{lsm-tree} based storage engine is widely adopted in modern cloud storage and database storage due to its write friendly storage structure and simple concurrency
control to support write intensive scenarios which  includes  enterprise storage servers and online transactions.
The idea of LSM-tree is to first buffer key-value(KV) writes in memory buffer and then flush these writes to disk files. Key-value pairs on disk files are sorted and disk files are structured as multiple level where lower level files contain most recent writes. Key-value pairs in lower 
level are gradually moved towards higher level via compaction process.
Typical LSM-tree key-value storage engines are RocksDB \cite{rocksdb}, Titan \cite{titan}, XStore \cite{xstore}, LevelDB \cite{leveldb} and TerakDB \cite{terakdb}.

Standard LSM-tree based storage engine that stores keys and value in LSM-tree suffers from 
write amplification because of repeated write of keys and values during compaction.
Such write amplification can reach a factor of at least 50x \cite{wisckey}. 
And read amplification also increases as LSM-tree grows in size because multiple disk accesses 
are required to locate the target key across multiple levels of disk levels.
Write amplification increases significantly when value get large.

Key-value separation is introduced by Wisckey \cite{wisckey}  to reduce write amplification from recurring large 
value write during compaction. The main idea of KV separation is to store large value in separate value files and store keys and pointers to value in LSM-tree. Individual garbage collection is initiated to reclaim storage space by checking validity of values , writing back value to new value files and writing KV pair that contains keys and latest value pointers to LSM-tree if necessary. 
This design helps to lower write amplification because it avoids repeated write of large value during compaction process and it's useful for workloads whose average value size of KV pairs are large. This design comes with cost of 
sacrificing scan performance for the reason that values are stored in value files out of order.

Current design and implementation of garbage collection process for KV separation  struggles to achieve low write 
amplification and low space amplification at the same time because current garbage collection 
process is static parameter based. Typically, garbage collection process is triggered when estimated or measured garbage ratio in value files exceeds some predefined threshold or when value files exceed defined time-to-live threshold. If garbage ratio for triggering garbage collection is low then we can get low space amplification with high write amplification. Otherwise, high garbage ratio for garbage collection triggering can cause low write amplification but with high space amplification which means it wastes storage space to store obsolete values.

We argue that the root cause for why current GC solution for KV separation struggles to balance low write amplification and low space amplification at
the same time is that key lifetime is not known in advance. 
Key lifetime is defined as duration from the time when the key is first initially written to the time it is 
rewritten which means previous write of key is invalidated.   
Despite that the idea of leveraging lifetime of blocks or objects to improve garbage collection efficiency has been studied in SSD \cite{pcstream} and file system \cite{fstream} it has not been studied how this idea can be applied to garbage collection in KV separation in LSM-tree.

This paper introduces DumpKV, an intelligent LSM-tree KV separation store that conducts lifetime aware garbage collection tailored for update intensive workloads.
The primary concept of DumpKV revolves around considering the lifetime of keys. It adaptively monitors the distribution of workload lifetimes and employs a machine learning model to understand this distribution. The model then provides lifetime predictions for each key. Based on specific lifetime thresholds, these values are subsequently placed into a value file. This approach ensures a more efficient and adaptive key-value storage system.
So  values that are expected to be invalidated at the same time can be garbage collected together which increase 
garbage collection efficiency. Thus low write amplification and space amplification can be 
achieved at the same time.
Specifically, DumpKV organizes value files into short and long lifetime categories whose value is dynamically adjusted based on lifetime distribution monitoring. Values are first written to short lifetime value life 
. During garbage collection process a binary classification model gives remaining lifetime prediction for KV pairs that are still valid based on past write information and then each value are written to value file with specified lifetime category. Garbage collection is initiated when the estimated time-to-live (TTL) of a value file is reached.

There are several challenges to be solved in order for DumpKV to incorporate model into LSM-tree KV separation store to achieve good garbage collection efficiency without degrading system read and write performance .

First, effective feature engineering are required for model to give accurate lifetime prediction. To generate effective features DumpKV uses past write information of keys and generate exponentially decayed write count that captures short term and long term write access pattern as features to do per key level lifetime prediction.

Second, it's required to determine short and long lifetime threshold properly. To detect short and long lifetime boundary for data labelling and for garbage collection trigger threshold DumpkV collects lifetime of keys
continuously and find the best short and long lifetime threshold based on cumulative distribution function  periodically.

Third, overhead of feature collection and model prediction should bring minimum inteference to system read and write performance.
To reduce overhead of data collection and model prediction DumpKV doing features generation during L0-L1 compaction and doing model prediction during garbage collection in background process. The main idea to reduce feature collection and model prediction overhead is to do them in low priority background process.
DumpKV also leverages the fact the LSM-tree is small enough under KV separation architecture  so read overhead of LSM-tree is relatively small.

Last but not least, DumpKV needs to handle dynamically changing workload and give good prediction of key lifetime in an adaptive way. Previous work that uses offline trained model in LSM-tree KV store is not impractical in pattern changing  workload .DumpKV solves this problem by doing model retraining periodically and adjusted training dataset 
generation in an adaptive way.

To the best of our knowledge, this is the first work that introduces learning method to help with garbage collection in KV separation in LSM-tree.
Contributions of this paper are summarized as follows.
\begin{itemize}
\item We propose DumpKV, an intelligent KV separation store architecture which learns lifetime distribution and gives per-key level lifetime prediction during garbage collection to achieve low write amplification and space amplification at the same time. 
\item We propose effective features engineering techniques solely based on past write access information and  feature persistence to help model do training and prediction.
\item We propose lifetime aware value file storage structure and  garbage collection process to effectively relocate  KV pairs of value files that are still valid. 
\item We implement DumpKV prototype atop of RocksDB, an open-source KV store that is popular in community and is widely adopted. Experimental results show that DumpKV reduces total write size ranging from 38\% to 73\% with slight extra storage space for feature data.

\end{itemize}

\section{Background }
In this section we will first give introduction to LSM-tree store architecture. We then give introduction to
key-value separation technique that aims to reduce write amplification for large value storage. And then we discuss
about current garbage collection design and implementations in key value separation in LSM-tree storage engine and gives strengths and weaknesses of different garbage collection approach. Finally, we talk about current learned based research approach in storage system that predicts lifespan of storage object.

\subsection{LSM-tree key value store}

\begin{figure}
    
    \includeinkscape[scale=0.20,pretex=\fontsize{20pt}{20pt}]{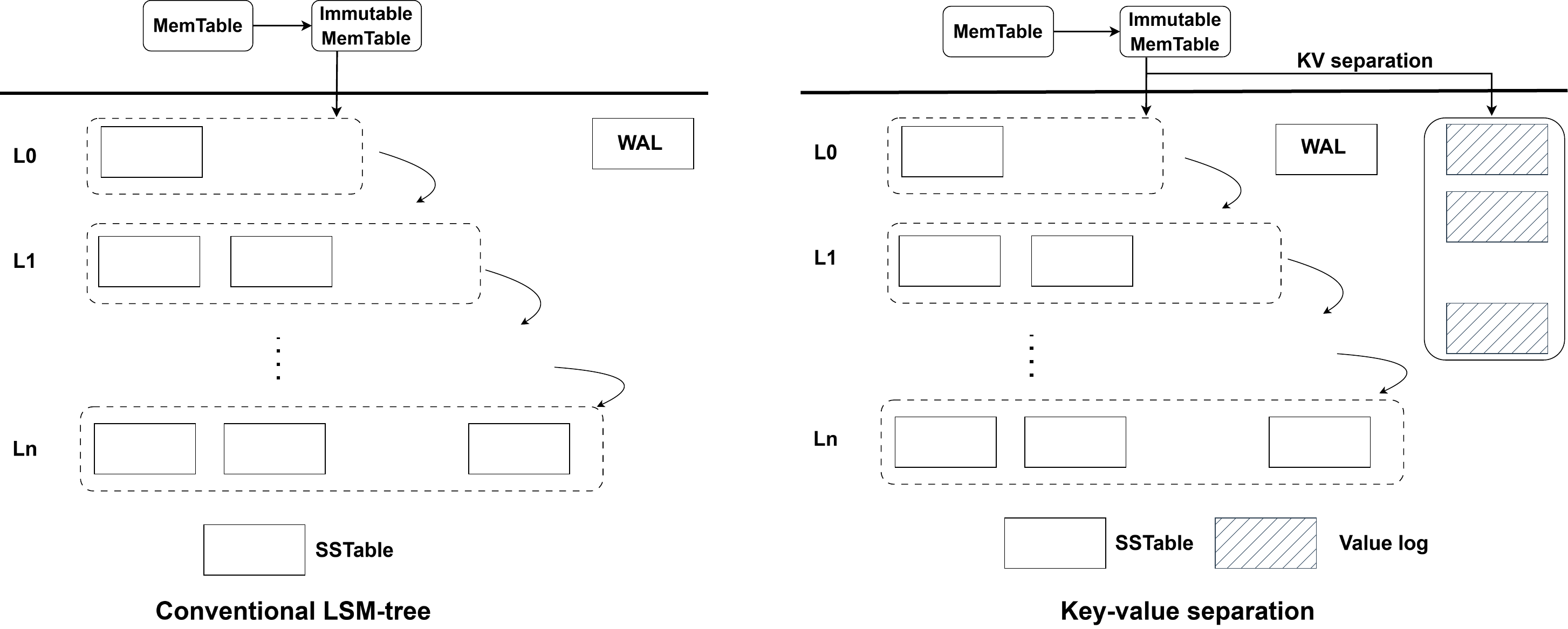}
    \caption{LSM-tree and KV separation}
    \label{fig:lsm-tree}
\end{figure}

LSM-tree key value store is a write friendly storage architecture. Fig \ref{fig:lsm-tree} depicts a simplified storage architecture of 
conventional LSM-tree key value store(eg, LeveldB and RocksDB).  LSM-tree key-value 
store is an append-only storage structure. KV pairs written by users are first buffered in memory part called MemTable. 
When MemTable reaches size threshold it's converted into Immutable MemTable which means there will be no extra writes to this Immutable MemTable and a new MemTable is created in memory to accept new KV pairs. Then Immutable MemTable is flushed to a disk file called SSTable . LSM-tree key-value sotre organizes SSTables in n+1 levels, denoted by $L0$, $L1$, ... , $Ln$  (from lowest to highest level) in disk. Capacity of $Li$ is configured as a multiple (typically 10x) of that in $Li-1$ (where  1  <= i <=n ).

Each level except $L0$ is fully sorted which means there is no overlap between SSTables. 
When $Li$ reaches size limit a compaction process is started to merge key value pairs from $Li$ to $Li+1$($0 \leq i \leq n-1$) .
Compaction process first picks candidate SSTables from $Li$ and then finds the overlapping
SSTables in $Li+1$ to be candidates in compaction. 
Then it sorts all key value pairs from candidate SSTables and writes valid key values pairs to newly created SSTables in $Li+1$. 
Finally input candidate SSTables are deleted and the compaction process is done. This compaction process incurs extra read and write which increases  write amplification.
Prior studies show that write amplification of conventional LSM-tree key value store can reach up to 50x write amplification.

To perform read process to do key lookup, a conventional LSM-tree KV store first search query key in MemTable, and if it does not find the query key then it performs another search in Immutable MemTable if there is any.
If the query key does not exist in MemTable and Immutable MemTable, LSM-tree KV store then searches in each level of LSM-tree starting from L0 to higher levels.
In $L0$, LSM-tree KV store searches all SSTables. In levels between $L1$ and $Ln$ LSM-tree KV store does binary search to locate the candidate SSTable whose key range overlaps with lookup key.

\subsection{Key-value separation and garbage collection}
Large size values are commonly found and contribute a considerable amount of traffic in real word KV workloads. Recent study from Facebook shows that value size of KV pairs from social graph data can reach as large as 1KB. Another example is that TiDB, a
transactional database built on top of KV storage engine, maps each row in table into a KV pair whose value size can reach hundreds of kilobytes.

To reduce write amplification caused by repeated large value write during compaction, Wisckey proposes KV separation architecture. 
Fig \ref{fig:lsm-tree} shows storage structure introduced by Wisckey.
Specifically, Wisckey puts values in separation value file and puts keys and index pointer
that points to position of value in value file as value in LSM-tree. This index pointer has much smaller fixed size compared to large value. Large values are not rewritten over and over again 
during compaction so compaction overhead and write amplification is reduced. 
Key-value separation also significantly reduces the LSM-tree size which reduces read amplification and benefits point query performance.
To do key lookup in KV separation, Wisckey first does a lookup in LSM-tree to locate KV pair in LSM-tree, and then it reads actual value of key from value file with index pointer value as reference.

 Garbage collection is required to reclaim storage space occupied by obsolete values in value files. 2 scan old value files and write back valid values to new value file. 
Several KV separation designs and implementations are introduced to trade-off between read and write performance brought by garbage collection \cite{wisckey, terakdb, diffkv}. They  mainly differs in value file storage structure, trigger condition of garbage collection process and victim value file selection strategy.

Wisckey\cite{wisckey} maintains circular value log(vLog) and inserts new values to the beginning of vLog and oldest values are at the end of vLog .
Wisckey initiates garbage collection job when vLog reaches size litmit and reads chunk of 
  values at the end of vLog and checks validity. Then it writes valid values to the head of vLog  and writes new value index pointer to LSM-tree. This value index pointer write back interferes foreground user write throughput.
Titan \cite{titan} adopts similar garbage collection policy as Wisckey but with different garbage estimation implementation. It triggers garbage collection when estimated garbage ratio of
value files reaches threshold\cite{diffkv}.
HashKV \cite{hashkv} proposes design on top of KV separation to uses hash-based data grouping to map values whose keys share same hash index to the same 
store segment so obsolete values in the same segment can be directly discarded because latest 
version of each key must reside at the end of the group.
But HashKV still has to iterate through all segment groups for garbage collection, a process that may not be efficient for bulk load scenarios or less skewed workloads. 
Moreover, HashKV still needs to write back 
valid value pointer to lsm-tree to update new value position which degrades foreground write 
performance. 

RocksDB\cite{rocksdb} develops its own KV separation design and implementation and integrates garbage collection as part of compaction process which is different from Wisckey.
Specifically, during compaction RocksDB fetches value from value files whose file numbers are below specified age cutoff value file number for each KV pairs in SSTables and writes values to new value files.
There are two benefits from this method. The first one is that there is no foreground write to Memtable to update value index pointer. Instead value index pointer update is done during the generation of new SSTable in the background so interference of garbage collection is limited. 
The second one is that  there is no read operation to query the validity of key. 
Keys and their values are dropped if there newer version of them in input SSTables. 
However, this skipped LSM-tree query step to check validity of key in the whole LSM-tree makes it likely that obsolete keys and values not discarded during garbage collection process because there might be newer version of obsolete keys that are the lower level of LSM-tree that are not involved in upper level garbage collection process.

TerakDB \cite{terakdb} removes new value pointer write-back by storing values in separate value file called v-SSTable during garbage collection. 
TerakDB is different from Wisckey in terms of value file storage structure and value pointer index format. 
Specifically, TerakDB stores file number in value index pointer only instead of file number and offset of value in value file. Values in value files are stored in special type of SSTable called v-SSTable in sorted order.
TerakDB maintains inheritance map between v-SSTable so key lookup can search this inheritance map and find the latest v-SSTable that contains the value. 

A fair amount of work has been done to reduce write amplification caused by garbage collection of
log-structured storage that apply lifetime prediction idea for data block \cite{pcstream, fstream, autostream}. SepBIT  \cite{sepbit} uses latest update interval and age of a block to estimate the invalidation time and assign each data block to hot or cold group. MiDAS \cite{midas} employs analytical models to minimize garbage collection overhead by taking update interval, frequency and age of block into account to give more precise invalidation time for hot and cold blocks and adjusting the number of groups and sizes according to I/O patterns to minimize the movement of data blocks.    

\subsection{Learning-based prediction in storage system }
A fair amount of work has been done to apply machine learning model do prediction for various parts in various aspects of storage system stack to benefit read/write rate. Typical use cases are increasing cache hit rate or improving garbage collection efficiency.
Leaper \cite{leaper}uses model to predict which new blocks are hot after compaction and prefetches those new blocks into cache during compaction. 
LRB \cite{lrb} uses machine learning to approximate the Belady MIN alroithm to predict objects with furthest request and evict those objects from cache to increase cache hit rate.

One main challenge in the integration of machine learning into storage systems is the necessity for the model to impose minimal training and inference overhead on critical system performance metrics, such as read and write throughput. Current solutions can be broadly classified into two categories. The first approach involves constraining the model size and simplifying input features to reduce both the training and inference time\cite{linnos, baleen, leaper}.
For example, 
LinnOS \cite{linnos} applies light weight neural network with weight quantization and reformatting  feature integers into decimal digits to do binary classification and do SSD performance inferring  at a per I/O granularity to reduce I/O latency.
The second one is reducing model calling frequency or putting model inference in non-critical path \cite{learnedcpp}.
For example, LLAMA \cite{learnedcpp} uses hashing-based mechanism to identify contexts previously seen and only execute model inference if the lookup fails to amortize model executions overhead over the lifetime of a long-running server.

\section{Motivation }

\begin{figure}
    \begin{subfigure}[b]{0.22\textwidth}
        \resizebox{\linewidth}{!}{
            \includeinkscape[scale=0.44]{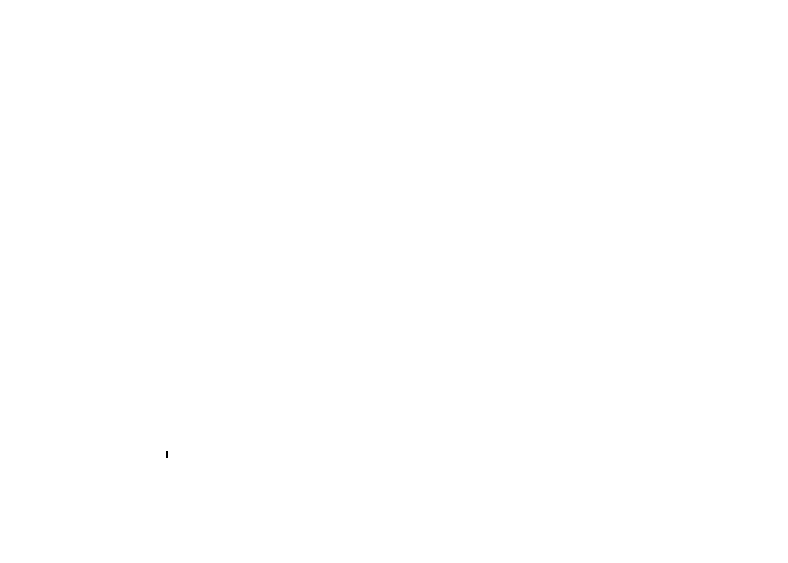}
        }
    \end{subfigure}
    \begin{subfigure}[b]{0.22\textwidth}
        \resizebox{\linewidth}{!}{
            \includeinkscape[scale=0.44]{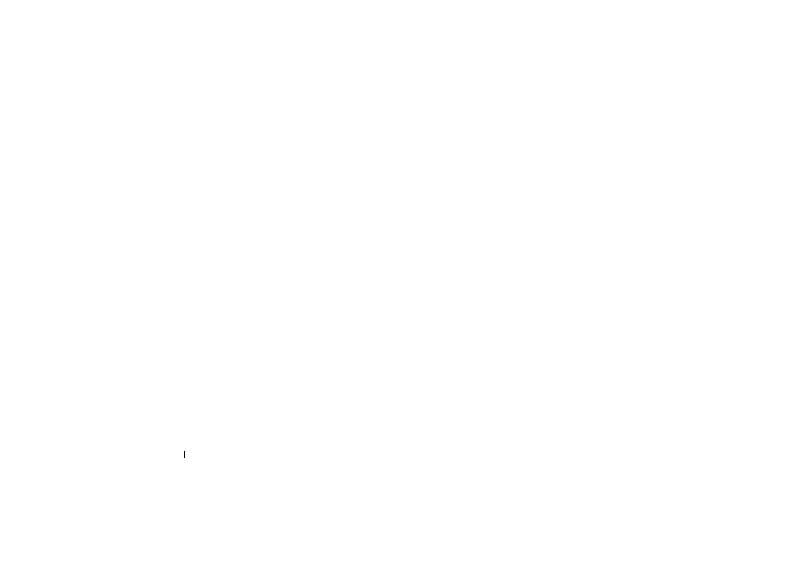}
        }
       \vspace{-11pt} 
    \end{subfigure}
    \caption{LSM-tree and KV separation}
    \label{fig:write-size-total-size}
\end{figure}

Despite that lifetime information is used for garbage collection in storage hardware such as SSD\cite{sepbit}, file system \cite{f2fs} and memory allocation \cite{learnedcpp}, it has not been studied how lifetime information can be leveraged to improve garbage collection efficiency and reduce write amplification in KV separation for LSM-tree.   

Besides, current garbage collection designs for KV separation suffer from balancing low write amplification and space amplification because they only take static parameters to trigger garbage collection.
Thus, current garbage collection implementation in key value separation in LSM-tree  either wastes trade off higher write amplification to have less space amplification or wastes extra storage space to maintain low write amplification.

A fair amount of works \cite{hashkv, diffkv} adopt techniques to differentiate hot and cold keys by recent write count, which is a simplified version of lifespan prediction of keys. This recent write count can be treated as a single feature for a binary classification task.
However, this differentiation technique still relies on threshold-based garbage collection triggers and struggles to catch lifetime patterns.
Thus, it cannot help achieve lower write amplification and space amplification at the same time when the workload is less skewed. This means that we have to take more features into account and give more fine granularity lifetime classification to keys in order to achieve lower write amplification and space amplification at the same time .

Fig \ref{fig:write-size-total-size} shows total write size and total size of three typical KV separation implmenetaion RocksDB, TerakDB and Tian under different garbage collection trigger ratio parameter settings. All 3 KV separation LSM-tree based stores show similar trend and lower garbage collection trigger ratio leads to lower total size but with higher write amplification.

If lifetime of each key can be known in advance then we can group keys with similar lifetime together and achieve higher garbage collection efficiency, thus write amplification and space amplification can be lowered at the same time. 
This motivates us to use the machine learning model to learn the distribution lifespan of keys to give a precise lifetime prediction to keys. 
There are several problems to be solved in order to achieve an effective and efficient learned LSM-tree KV separation store.

First,   effective features generation for keys for training and inference is required  without full stack application information.
Limited access information is available from the point of view of LSM-tree key-value store. 
Unlike PCStream\cite{autostream} that gathers call stack traces and program contexts from file system to do hot and cold data separation and 
LLAMA \cite{learnedcpp} which is a learned memory allocator for C++ that uses  rich call stack traces as input features to train neural network model to predict lifetime for memory objects there is no such data for a standalone key value store.
So it's required to do efficient feature engineering and generate features based on past write access information that can help model capture lifetime patterns.  

Second,  overhead of model training and inference should bring minimum impact to system  read and write performance.
Unlike user-facing machine learning application, the LSM-tree key value store is sensitive to read and write performance. 
It's critical to reduce overhead of model inference and training while maintaining good balance between write amplification, space amplification and write performance. 

Third, features of keys need to be stored persistently  and efficiently to be used for model training and inference.
Features of keys need to  avoid data loss and  they need to be quickly retrieved for model inference to predict lifetime of next write to the same key. 

Finally, the model needs to be updated online to adapt to the change and shift of the workload pattern.
User workload pattern can be dynamically changing so it's important to keep model updated to the latest lifetime distribution of keys which is critical to maintain good prediction accuracy.

\section{DumpKV Design}
In this section we present the design of DumpKV and illustrate how we address the problems mentioned in the last section. First, we give DumpKV system overview. Then we describe how features are generated for use in model prediction and how features are stored efficiently and effectively.
Next, we introduce how model training and inference works in DumpKV. And then we talk about how garbage collection works in DumpKV. Finally we discuss how crash consistency mechanism and implementaion details.

\subsection{System overview}

\begin{figure}[h]
    \centering
    \includeinkscape[scale=0.44]{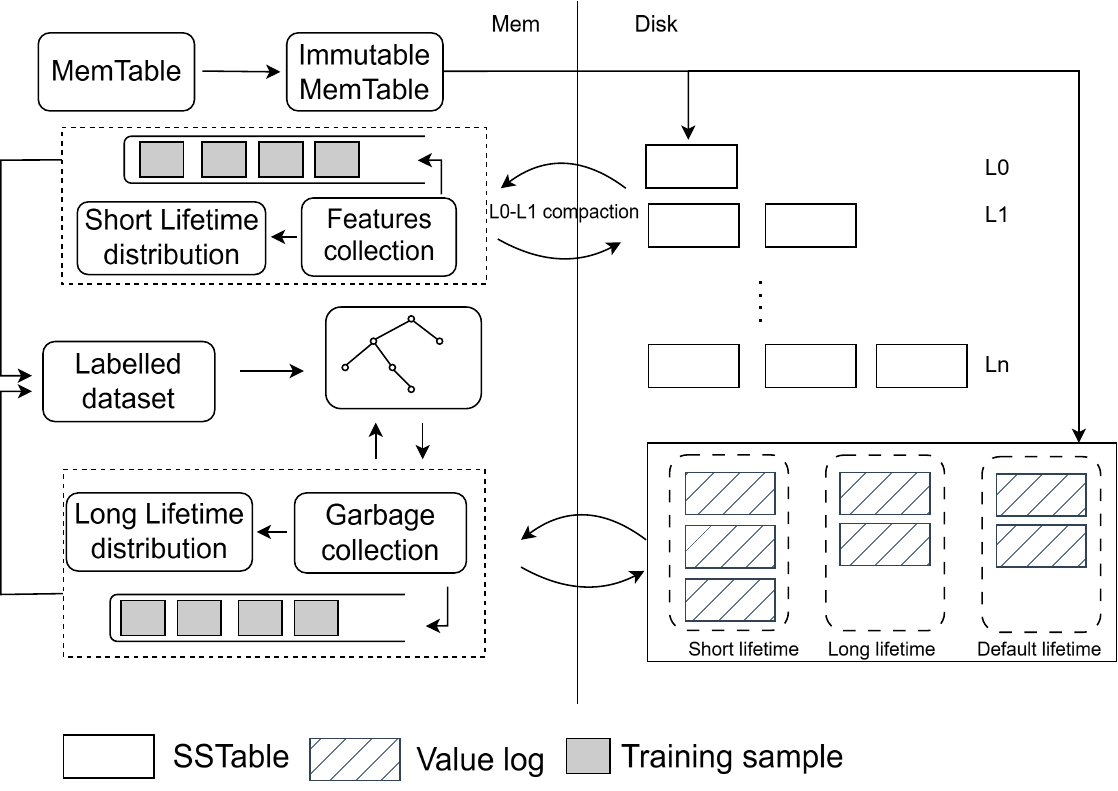}
    \caption{DumpKV architecture}
    \label{fig:dumpkv-arch}
\end{figure}

Fig. \ref{fig:dumpkv-arch} shows the system architecture of DumpKV. 
DumpKV is a key value separation design of LSM-tree that applies machine learning model to learn lifetime distribution of keys and give 
lifetime prediction to each key and puts value of key to value file with lifespan limit . 
Large values are first written to default lifetime value files whose lifetime threshold that is dynamically adjusted.
Features data collection and lifetime collection happens in $L0-L1$ compaction and garbage collection process. And model training is triggered periodically after training dataset accumulates enough training samples. 
Garbage collection is initiated once the value files have attained their anticipated lifespan, and the model gives remaining lifetime prediction for each KV pairs that are still valid.

To generate effective features with only past access information to help model give good remaining lifetime prediction and make this method applicable in
various usage scenarios DumpKV takes  past update intervals and 
exponentially decayed write counters that are calculated with constant time and space as features.   
To reduce overhead of model training and inference  to system write performance. DumpKV uses lightweight gradient boosting tree model 
to do training periodically in background thread and do model prediction  during garbage collection which happens at background as well.  
DumpKV puts KV pairs to value files with adaptive default lifetime generated in Flush process instead of calling model to predict remaining lifetime to balance write performance and write amplification.  Because it's very expensive to generate feature data and doing model inference during Flush which decreases system write performance.   

To store features generated effectively and efficiently for model training and inference and prevents data loss in case of system crash 
DumpKV stores features as part of value in the LSM-tree. Features of each key that are used for training and inference can be done by querying LSM-tree which brings small overhead because of small LSM-tree size under KV separation architecture.

To achieve online learning and continuous learning DumpKV does feature data generation and model training periodically to help model  adapt to  workload pattern shift.

DumpKV performs binary classification training for two primary reasons:
The first one is training a multi-classification task can be challenging, and there’s a risk of misclassifying adjacent lifetime samples \cite{linnos}.
The second one is that the presence of multiple categories in the lifetime of value files increases the total number of these files, which can negatively impact read and write performance. Consequently, the scan operation would need to process more files to retrieve all values.

\subsection{\textbf{Features and storage}}

\subsubsection{\textbf{Features}} 
DumpKV uses dynamic and static features together to train model.

\textbf{Deltas.}
 $Delta$ is defined as interval between consecutive write of the same key.
 $Delta_0$ is the elapsing time since the latest previous writes.
$Delta_1$ indicates the time interval between previous two writes. 
We choose to use key sequence  as time unit instead of real time such as microsecond or second because of following two reasons.
First, Time movement is driven by new incoming writes rather than clock time, which is a more logical trigger for initiating a garbage collection job.
Second, key sequence is used as unique timestamp in current LSM-tree store engine so it is directly available. 

DumpKV uses up to 32 $Delta$ for a key and $Delta_i$ where $i >= 32$ is discarded. 
To maintain numerical stability of training data DumpKV maps $Delta$ value to index of exponentially increasing intervals which are 1M, 2M, 4M, 8M, etc.
For example, $Delta$ with 3M value is mapped to 2 in training data.
We uses past update interval as features because it captures the past write access information of a key .

$Delta$ values remain constant over time and are unaffected by the arrival of new writes for the same key. This characteristic simplifies storage and minimizes the need for frequent updates of feature values. It should be noted, however, that the storage space required for the $Delta$ varies for each key.
For example, key A with 10 past writes has 9 $Delta$,
while key B with 2 past writes only has 1 $Delta$ that occupies storage space. This helps save space for feature storage with less frequently updated keys.

\begin{table}[h]
    \centering
    \caption{List of notations used in section below}
    \begin{tabular}{p{.25\linewidth}|p{.6\linewidth}}
    \hline
       Notation  &  Description \\ \hline
        $H_{s}(x)$, $H_{l}(x)$ & Lifetime distribution histogram for short and long lifetime value files \\  \hline
        $l_{s}$, $l_{l}$ &  TTL for short and lifetime value files \\ \hline
        $l_{d}$ &  TTL for default lifetime value files generated in Flush \\ \hline
        $s\_idx$, $l\_idx$ & Lower bound index for $l_{s}$  and $l_{l}$ in $EDWC_i$ windows \\ \hline
        $sp_0$, $lp_0$ & Base percentage value for dynamic short and long  lifetime for value files \\ \hline
        $sp_1$, $lp_1$ & Extra percentage value for dynamic short and long lifetime for value files \\ \hline
        $ini\_{sp}_0$, $ini\_{lp}_0$ &  Initial percentage value for base percentage value for $sp_0$ and $lp_0$  \\ \hline
        $ini\_{sp}_1$, $ini\_{lp}_1$ & Initial percentage value for extra percentage value for $sp_1$ and $lp_1$ \\ \hline
        $\alpha_{s}$, $\alpha_{l}$ &  Steepness factor for  transformation function  \\ \hline
        $\beta_0$ , $\beta_1$ & Midpoint shift for transformation function  \\  \hline
        
    \end{tabular}
    \label{tab:my_label}
\end{table}
 
\textbf{Exponentially decayed write counters(EDWCs)} 
\label{label:edwc}
DumpKV applies EDWCs as another feature to capture update frequency information.
EDWC tracks the write access count in a specific time window. 
When a new write request arrives, $EDWC_i$ is updated according to following expression
\begin{equation}
\label{eq:edwc}
    EDWC_i = 1 + EDWC_i \times 2^{-Delta_0 / 2^{19+i}} 
\end{equation}

Each $EDWC_i$ is not updated until a new write request of key arrives which is different than that $Delta_0$ is updated each time it's used for model prediction.
$EDWC_i$ with larger value i cover longer time window. For example, a key that is updated frequently 10M timestamp ago but is rarely updated now would have a low $EDWC_1$ but still have a large $EDWC_{4}$.
So Keys with few write access recently could still have high write access count in the long term.

$EDWC$ is motivated by prior work in block storage caching \cite{lrfu}, cache prediction  \cite{lrb} and video popularity prediction \cite{chess}. The idea is to use $EDWC$ to track long term trend of popularity by approximating decay rate of  object popularities.
Write access pattern of keys are captured with multiple $EDWC$ with different decayed constants that represent different time windows.

\textbf{Additional optional features} Besides dynamic features that are updated when each new write  arrives, other static features 
can be used to help model determine the lifetime of keys. For example, the size of the value and the type of key that is related to the application can be utilized. These features can be useful for an application-specific scenario, and they require little overhead to store.

\subsubsection{\textbf{Feature storage} }
\begin{figure}
    \centering
    \includeinkscape[scale=0.44]{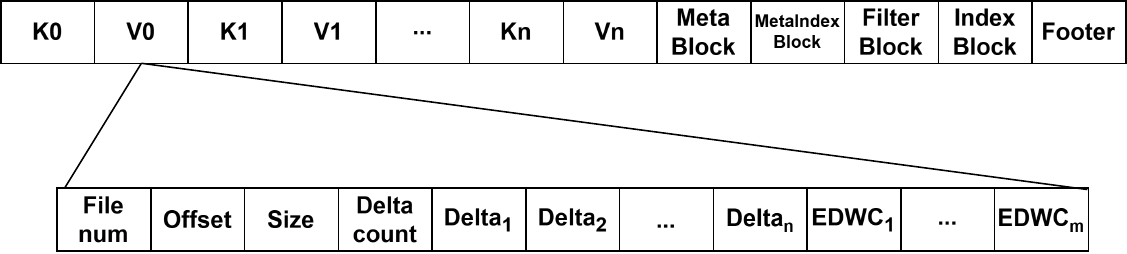}
    \caption{Feature storage format in LSM-tree}
    \label{fig:feature-format}
\end{figure}
Fig \ref{fig:feature-format} shows feature storage format. 
Value part in SSTable in LSM-tree first stores value file number, value offset and value size in corresponding value file. After these three fields DumpKV stores feature data to be used for future model prediction. $Delta Count$ is used to indicated how many $Delta$ are stored in the value part and it only takes one byte storage space. Following that are actual value for $Delta$. Following that is $EDWC_i$ data for current key.
DumpKV stores up to 32 past distances delta and 10 EDWC for each key that has at least one previous writes. 

Table \ref{table:feature-storage-size} shows feature data storage size for keys with different accumulated write count.
Only 1 byte is required for  keys that are first written to storage engine and  have no past writes  because no past distances and EDWC features are stored in LSM-tree. 
Maximum storage space for feature data is 296 bytes which is affordable given that value size is usually large in KV separation architecture and it's rare to have keys with more than 32 past writes take up majority of overall write requests.
DumpKV uses varint encoding to save more storage space.
\begin{table}[h]
\centering
\begin{tabular}{c|c|c|c|c|c|c|c}
\hline
\textbf{\# of past writes } & 0 & 1 & 2 & 3 & 4-12 & 13-32 & > 32  \\
\hline
\textbf{Feature size} &  1 & 49 & 57 & 65 & <127 & <297 & 297\\
\hline
\end{tabular}
\caption{Feature storage overhead for keys with different number of past write requests.}
\label{table:feature-storage-size}
\end{table}

\subsection{\textbf{Training data collection }}
DumpKV maintains two training sample collection queues to collect training sample from $L0-L1$ compaction and garbage collection separately.
A separate consumer thread is then responsible of gathering samples from these two queues, doing data labelling and building training dataset .

The rationale behind maintaining two separate queues for collecting training samples is as follows:
True lifetime label for past write of the same key  can be  obtained during $L0-L1$ compaction. This queue helps model to learn lifetime distribution for keys that have at least two writes.
To help model to pay attention to one time write keys and keys that expect to have long lifetime DumpKV maintains another queue to collect training data from garbage collection process.

In the process of $L0$-$L1$ compaction, for each key that is iterated, DumpKV performs a search operation in levels $Li$ 
 where  $1 \leq i$
 to identify any previous write of the same key. The lifetime of the previous write is calculated by subtracting the sequence number of the current write from the sequence number of the previous write. Subsequently, the feature data of the previous write is extracted from the value part in the LSM-tree and is appended to the $L0-L1$ compaction queue, along with its lifetime. 
 Then all $EDWC_i$ of previous write of the same key is updated according to Eq.\ref{eq:edwc} and they are written to value part in LSM-tree together with latest $Delta$ feature with feature storage format as shown in \ref{fig:feature-format} for future feature collection , model training and calling. 
 
To avoid frequent written keys are added to training dataset during $L0-L1$ compaction process DumpKV adds training sample to $L0-L1$ compaction sample collection queue with probability 
\begin{equation}
p = EDWC_{s\_idx} / EDWC_{l\_idx}
\end{equation}
The variables ${s\_idx}$ and ${l\_idx}$ 
 represent the indices of the EDWC window, which correspond to the upper bounds of the current short lifetime point and long lifetime point, respectively.
 The size of the LSM-tree is inherently small under KV separation architecture, which results in a minimal overhead when performing point queries within the LSM-tree, without the need to read values from the value file. This efficiency allows for the majority of the LSM-tree to be stored in the block cache.

 During the garbage collection process, DumpKV calculates the elapsed lifetime for each valid key. This is achieved by subtracting the sequence number of the key from the current timestamp sequence number. Subsequently, the feature data of this key is extracted from the value part. This data is then appended to the garbage collection training sample collection queue, along with the elapsed lifetime is included as a speculated lifetime label.
 The probability of this action is equivalent to the current garbage collection valid ratio.

Please see section \ref{sec:data_labelling} for details about how data labelling is conducted.

\subsubsection{\textbf{Short and long lifetime point}}
During the training data collection process, lifetime distribution monitoring is conducted to identify optimal short and long lifetime points. These two points are subsequently utilized for data labeling in collected training samples.

DumpKV employs two histograms namely short lifetime histogram and long lifetime histogram to capture short and long lifetime points, a strategy rooted in the observation that lifetime information obtained during $L0-L1$ compaction accurately represents the true lifetime distribution for keys subjected to at least two writes.  Conversely, lifetime data collected during garbage collection fails to provide an accurate representation of the ground truth lifetime for keys that are long-lived, particularly those that have been written only once. This discrepancy underscores the rationale for this dual histogram approach.
Short lifetime histogram is updated for each ground truth lifetime obtained from $L0-L1$ compaction training samples queue. And long lifetime histogram is updated for each speculated lifetime derived from 
garbage collection training samples queue. 
The histogram employs a range-based approach to obtain distribution data, and utilizes a lock-free method for each update, thereby ensuring that the overhead of this lifetime distribution monitoring remains minimal.

And then short and long lifetime points $l_{s}$ and $l_{l}$ is calculated periodically based on current garbage collection ratio with following expression

\begin{equation}
\label{eq:short-lifetime}
\begin{cases}
 sp_0&=  ini\_{sp}_0 *  \sigma( \alpha_s * (1.0 - gc\_ratio - \beta_0))  \\
 sp_1&= ini\_{sp}_1 * \sigma( \alpha_s * (1.0 - gc\_ratio - \beta_1))  \\
 l_s&=  H_s(sp_0 + sp_1)
\end{cases}
\end{equation}

\begin{equation}
\label{eq:long-lifetime}
\begin{cases}
 lp_0&=  ini\_{lp}_0 *  \sigma( \alpha_l * (1.0 - gc\_ratio - \beta_0))  \\
 lp_1&= ini\_{lp}_1 * \sigma( \alpha_l * (1.0 - gc\_ratio - \beta_1))  \\
 l_l&=  H_l(lp_0 + lp_1)
\end{cases}
\end{equation}

where 
\begin{equation}
    \sigma(x) = 1 / (1 + e^{-x})
\end{equation}

DumpKV calculates $sp_0$ and $lp_0$ to get a base percentage value for short and long lifetime value files which ensures that lower bound of lifetime is close to 0 if $gc\_{ratio}$ is close to 1. Variables $sp_1$ and  $lp_1$ is more sensitive to low $gc\_ratio$ because $\beta_1$ is higher than $\beta_0$ and are used to be added with $sp_0$ and $sp_1$ to get final percentage value to get a proper lifetime value from lifetime histograms $H_s$ and $H_l$.
Variables $\alpha_s$ and $\alpha_l$ control magnitude change for lifetime point. Variables $\beta_0$ and $\beta_1$ control steepness for lifetime value change. $H_s$ and $H_l$ are histogram functions which return value given  specified lifetime distribution percentage. 
The rationale for these two expression is to assign a higher lifetime value point when the $gc\_{ratio}$ is low, thereby increasing garbage collection efficiency, and vice versa.
 Adaptive default lifetime for value files generated in flush process is then generated by taking current $gc\_{ratio}$ into account with following expression

\begin{equation}
   l_d = l_s + (l_l - l_s) * \sigma(1.0 - gc\_ratio - \beta_1)
\end{equation}
The idea is that default TTL for value files generated in Flush moves towards $l_{l}$ when $gc\_{ratio}$ is low and vice versa.

To mitigate the issue of data imbalance in the training dataset, DumpKV establishes an equal sample count threshold for each category of samples coming from $L0-L1$ compaction sample queue and garbage collection sample queue.

\subsubsection{\textbf{Data labelling}} 
\label{sec:data_labelling}

Data labelling is required to assign a lifetime label for each training sample to train the model. 
DumpKV does binary classification task and gives binary labels to indicate that whether 
a key will have short remaining lifetime or long remaining lifetime.

DumpKV applies different data labeling strategies for samples coming from $L0-L1$ compaction training sample queue and those coming from the garbage collection training sample queue.
For training samples that come from $L0-L1$ compaction queue DumpKV   excludes samples whose ground truth lifetime is shorter than default lifetime $l_{d}$ because those keys with lifetime that is shorter 
than $l_{d}$ are dropped during first garbage collection anyway so there is no need to pay attention to these keys.
Subsequently,  a sample $k$ is labelled with following expression 

  \begin{equation}
    \label{eq:compaction_label}
    label_k =
    \begin{cases}
      1 & \text{if } Delta_0 - l_{d} > l_{s} \\
      0 & \mbox{otherwise}
    \end{cases}
  \end{equation}

Equation \eqref{eq:compaction_label} shows that a sample is labelled as 1 i.e. long remaining lifetime if 
duration between last two adjacent write requests is greater than the sum of short lifetime threshold and default lifetime threshold for value files.

The process of labeling samples from the garbage collection sample queue presents a challenge due to the unknown ground truth lifetime. Despite this, it is imperative to include keys in the training dataset that have either a long lifespan or are written only once. This inclusion allows the model to predict a longer remaining lifetime for these keys during the garbage collection process.
DumpKV introduces a simple but effective heuristic rule based labelling method.
It assigns a label of 1, indicating a long remaining lifetime, to samples that are written only once.

The following expression is used to assign label for sample $g$ with at least two past writes from garbage collection training sample queue
  \begin{equation}
    \label{eq:compaction_label}
    label_g =
    \begin{cases}
      1 & \text{if } EDWC_{s\_idx} > 1.0 \\
      0 & \mbox{otherwise}
    \end{cases}
  \end{equation}

The rationale is that the model is advised to deem this key as inactive  if keys have fewer than one write operation in the $l_{s}$ window, implying a long remaining lifetime.

\subsection{\textbf{Model training }}
 
\hspace{5pt} \textbf{Model}
DumpKV uses Gradient Boosting Machine(GBM)\cite{gbm,lightgbm} as model to help train and do prediction of lifespan of keys. 
GBM is a machine learning model that is lightweight and is highly efficient on CPUs and widely used in many online machine learning application with tabular data\cite{resource-central, gbm-bing, gbm-bing-predict-clicks, gbm-click-through}, which is suitable for performance sensitive storage application. 
Other models such as linear regression, support-vector machine, logistic regression, two layer neural network struggle to outperform GBM on similar task \cite{lrb}.
Besides, GBM can handle missing values in an efficient way and does not require feature normalization\cite{lrb, lightgbm}.

\textbf{Prediction target} 
The model’s prediction target is to determine whether the remaining lifetime of keys, which are still valid during garbage collection, is short or long. This transforms the training task into a binary classification task. Specifically, samples with a predicted remaining lifetime close to the short lifetime point are expected to yield a model output close to 0. Conversely, samples with a predicted remaining lifetime exceeding the short lifetime point are expected to yield a model output close to 1.

\textbf{Loss function}
DumpKV uses binary logloss as loss function to train GBM model.  Binary logloss is a typical loss function  used to train binary classification model.

\textbf{Model training }
Model training is started once training samples reaches threshold.  
Specifically, DumpKV creates a new GBM model and trains this new GBM model with new training dataset with specified loss function.  
After training is finished the newly trained GBM model replaces previous  model and memory of previous model is deallocated. 
Because this training process is done in separate thread and does not block flush operation or compaction process it brings small overhead to performance of storage engine.

\subsection{\textbf{Garbage collection}}
\begin{figure}[t]
    \centering
    \includeinkscape[scale=0.44,pretex=\fontsize{20pt}{20pt}]{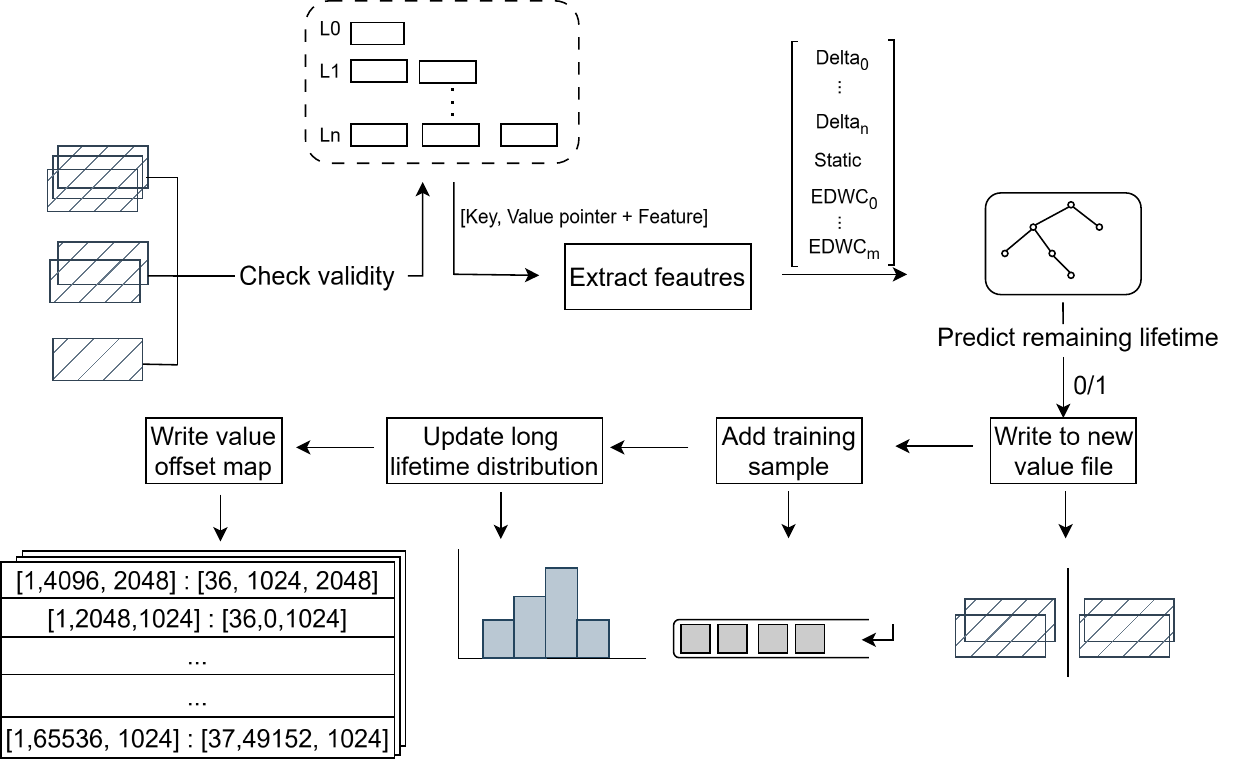}
    \caption{DumpKV garbage collection process }
    \label{fig:dumpkv-gc-process}
\end{figure}

Garbage collection job is triggered once lifetime TTL of value files is reached.
Fig \ref{fig:dumpkv-gc-process} show garbage collection process in DumpKV.
During garbage collection job DumpKV checks validity of each value by doing search in LSM-tree. This search process can be fast by reading block from block cache because LSM-tree is small so that majority of LSM-tree can be stored in block cache. If the value is invalidated and  obsolete then this value is discarded and no more extra steps are taken.

Otherwise, if the value is still valid then DumpKV reads feature data of this KV pair in LSM-tree during validity checking which requires no extra read, prepares  $Delta_0$  by subtracting current timestamp sequence number by sequence number of KV pair and updates each $EDWC_i$ with expression mentioned in \ref{label:edwc}. 
Then DumpKV calls GBM model to give remaining lifetime prediction with latest feature data of this value and decides to which lifetime category value file should this value written. 
After that DumpKV creates a training sample with current feature data and appends it to garbage collection training sample queue for generation of training dataset for next model. 
And then DumpKV updates long lifetime distribution histogram by adding $Delta_0$ value to it.

After garbage collection job is finished, DumpKV re-calculate default, short and long lifetime point $l_d$, $l_s$ and $l_l$ according to expression \ref{eq:short-lifetime} and \ref{eq:long-lifetime} based on latest $gc\_ratio$.
Note that value files with different lifetime threshold can be put into the same garbage collection job.

A value index map is created for each newly created value file during garbage collection.
For example in Fig\ref{fig:dumpkv-gc-process}, a value originally in value file:1 whose offset is 4096 and value size is 2048 is rewritten to value file:36 with offset 1024. 
This value index map is necessary because value index pointer that points old value  file in LSM-tree is unchanged during garbage collection job. 
Future read operation can get the value in latest value file by indexing this value index map. 
This index map is immutable after it's created and is stored in file to prevent data loss. Noted that because keys in value index 
map are value file number and offset so they are in increasing order which means that binary search can be done as well as hash 
search if this value index map is loaded in memory.
Value pointer index in LSM-tree is lazily updated to latest value by searching value index map when compaction is triggered. 
DumpKV deletes this value index map when no older value file numbers that could possibly depend on it.

\subsection{\textbf{Crash consistency}}
DumpKV maintains features crash consistency by storing features of keys in value pointer index in LSM-tree store engine,
Model crash consistency is achieved by storing GBM model parameters into file each time after model finish training. So model 
parameters can be restored next time when LSM-tree store engine restarts.
DumpKV doesn't persist training data into file so there is no need to worry about crash consistency for training dataset because 
training dataset is dynamically created in the running of LSM-tree engine. 
DumpKV handles crash consistency in garbage collection that is similar to compaction job. Since DumpKV doesn't write back new value 
pointer index to LSM-tree there is no need to worry about duplicate writes caused by garbage collection job like that in Wisckey.
If crash happens when new value file is created during garbage collection job then unfinished newly created value file and value index map file are discarded after DumpKV restarts.

\subsection{\textbf{Implementation details}}
We implement DumpKV on top of RocksDB v8.00. 
DumpKV checks whether each value file reaches lifetime limit after finish of each flush, compaction or garbage collection job.
DumpKV implements separate garbage collection module along with existing compaction and flush module. 
Training dataset size is set to 128k by default. 
Compressed sparse row(CSR) format is used for model training and inference because $Delta$ values for keys can be missing.

To serve read requests, DumpKV first does query in LSM-tree to fetch KV pair  which has potential invalid value pointer. To get latest value pointer for query key DumpKV searches value offset maps with initial value pointer in LSM-tree in an iterative way. 
Latest value pointer for query key is then obtained in the last value offset map. Then DumpKV reads value from value file with latest value pointer and returns value back to user.

\section{\textbf{Evaluation}}

In this section, we evaluate and compare DumpKV with three  state-of-the-art KV stores with KV separation design and implementation which are RocksDB, TerakDB and Titan. RocksDB has standard LSM-tree KV store implementation which is referred as RocksDB-std and KV separation store implementation which is referred as RocksDB in following evaluation part. 
Our goal is to answer following questions.
\begin{itemize}
\item  How is performance of DumpKV compared to other KV stores in terms of  write amplification, total size and throughput ? 
\item  What's the storage overhead of features ? 
\item  What is model calling overhead in garbage collection process and how does each feature contribute to model performance gain?
\item  How does dynamic lifetime adjustment of DumpKV work for workload with different skewness ? 

\item What's the performance of DumpKV compared to other KV stores under different varying value sizes? 
\item How does model prediction of DumpKV for remaining lifetime of keys impact write amplification and total size for workload with different skewness  ?
\end{itemize}

\FloatBarrier

  \begin{figure*}
        \resizebox{\linewidth}{!}{
            \includeinkscape[scale=0.40]{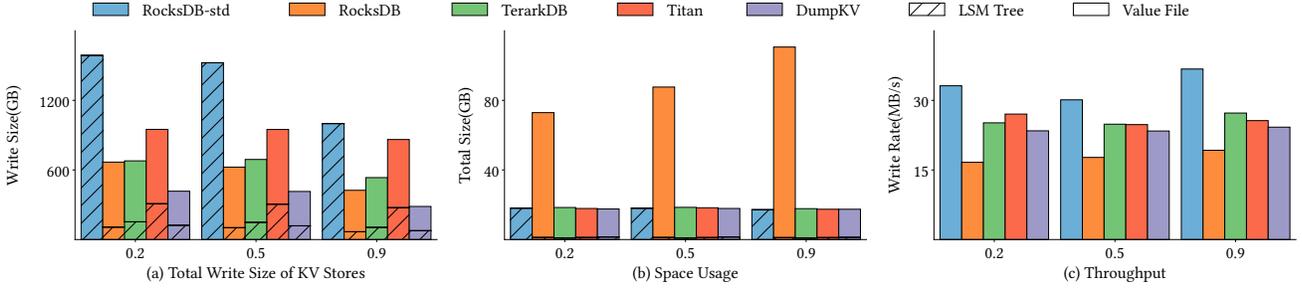}
        }
        \caption{Performance comparison results for different KV stores in terms of total write size, total storage space size and throughput. }
  \end{figure*}
\label{fig:diff-zipfian-50M}

\subsection{\textbf{Experiment Setup}}

  \begin{figure}[t]
  \centering
        \begin{subfigure}[t]{0.22\textwidth} 
               \raisebox{-5pt}{\includeinkscape[scale=0.28, pretex=\fontsize{5pt}{5pt}]{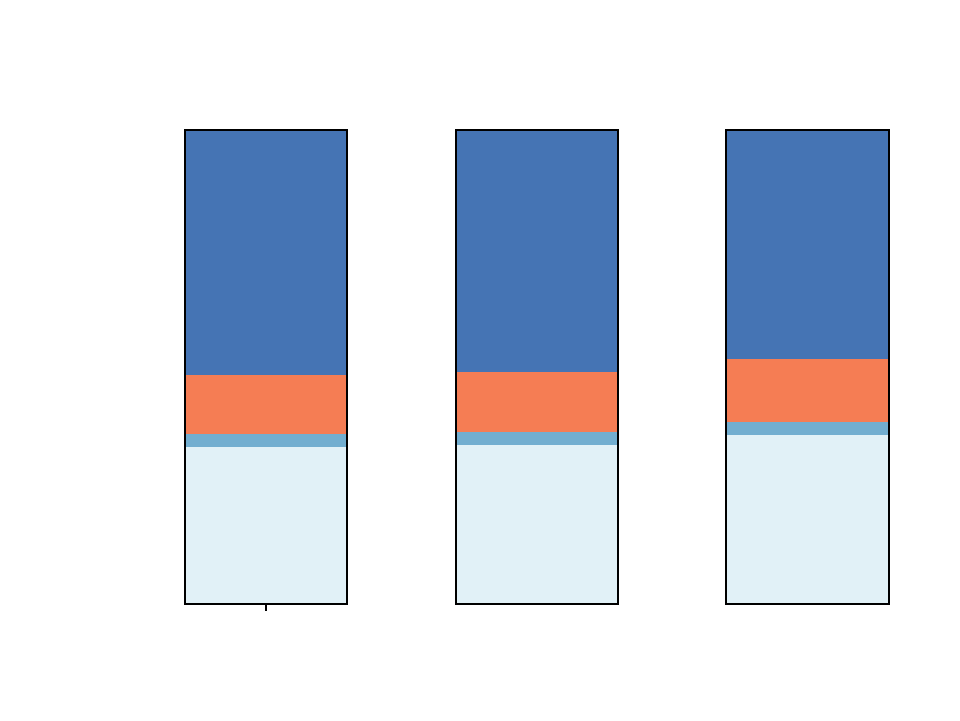} }
               \caption{(a) Model calling overhead }
            \label{fig:model_overhead_1}
        \end{subfigure}
        \hfill
        \begin{subfigure}[t]{0.22\textwidth} 
        \captionsetup{skip=10pt} 
      \includeinkscape[scale=0.25, pretex=\fontsize{5pt}{5pt}]{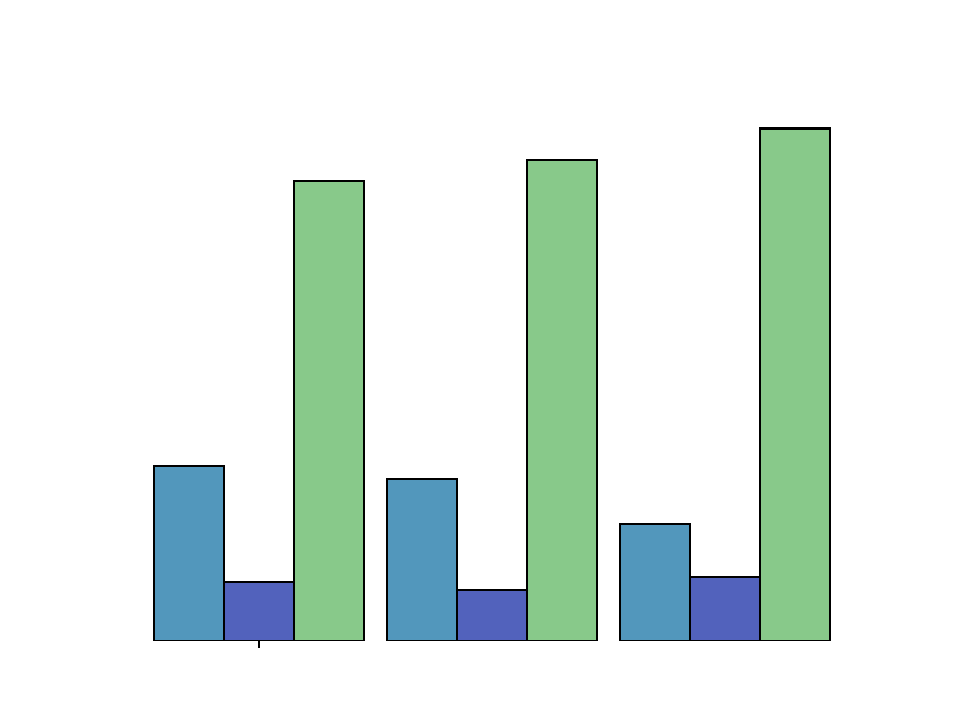}
      \caption{(b) Feature importance  }
      \label{fig:feature_importance}
        \end{subfigure}
   \caption{Model calling overhead for different skewness workload}
    \label{fig:model_overhead}
  \end{figure}

\textbf{Testbed} We run experiments on multiple machines equipped with 40-cores Intel Xeon Gold 6230N CPU, 196GB memory,  and Intel PEDME016T4 1.5TB SSD with Ubuntu 22.04 LTS .

We use YCSB\cite{ycsb, ycsbc} workload for testing. According to the actual experimental requirements, we generated zipfian distributed workloads with a variety of skewness parameters, including 0.2, 0.5, 0.9, etc.

\noindent \textbf{System configuration.} For all KV stores, MemTable size is set to 100MB, value file size is set to 256MB. SSTable file size is set to 32 * 10 * MemTable\_size / value\_size.  
For Titan, TerakDB, RocksDB and DumpKV the number of background garbage collection threads to 1 . Maximum number background compaction jobs is set to 16. 
Compression algorithms is disabled for all KV stores. Bloom filter is enable for all KV stores and number of bloom filter bits is set to 10 according to official tunning guideline. 
Block cache size is set to 16 GB and is used to cache blocks of SSTable only and cache for blocks of value files are disabled. 
For Titan and TerakDB, garbage collection triggering ratio is set to 0.2.
For RocksDB, we set \textit{blob\_garbage\_collection\_age\_cutoff} to 0.8   and  \textit{blob\_garbage\_collection\\ \_force\_threshold} to 0.2.

For DumpKV, $\alpha_s$ and $\alpha_l$ are set to 10. $\beta_0$ is set to 0.25 and $\beta_1$  is set to 0.75. For short lifetime point calculation, $ini\_{sp}_0$ is set to 60 and $ini\_{sp}_1$ is set to 40. 
For long lifetime point calculation, $ini\_{lp}_0$ is set to 80 and $ini\_{lp}_1$
is set to 20.

\subsection{Performance comparison}
We evaluate and compare performance of different KV stores under write intensive workloads. Specifically, we generated three workload data using YCSB with different skewness parameters which are 0.2, 0.5 and 0.9. Low skewness means that cold keys have more writes. Key size is set to 256 bytes and value size is set to 4KB. Each KV store is assumed to be empty initially. We then load 200GB data into each KV store separately. Requests are issued to each KV store as fast as possible to do stress test. Fig \ref{fig:diff-zipfian-50M} shows 

\noindent \textbf{ Total write size.} Compared to RocksDB-std, DumpKV reduces total write size by 71\%-73\%.  
RocksDB-std shows highest total write size because it repeatedly writes large values during compaction process. 
DumpKV shows lowest total write size among all KV stores which demonstrates that lifetime aware garbage collection can significantly reduces write amplification by grouping large values with similar lifetime, thus achieving efficient garbage collection.
Compared to RocksDB, DumpKV reduces total write size by 32\%-37\%.
Compared to TerakDB, DUmpKV reduces total write size by 38\%-46\%.
Compared to Titan, DumpKV reduces total write size by 56\%-66\%.
Titan shows second highest total write size because it repeatedly writes KV pairs with latest value pointer to LSM-tree during garbage collection.

\begin{table}[]
\caption{Training dataset size and model calling overhead for workloads with different skewness}
    \centering
\begin{tabular}{c|p{1cm}p{1cm}p{1cm}p{1cm}p{1cm}}
    \hline
     \centering  {\small Skewness }& {\small Trn count } & {\small Dataset size}  & {\small  Model  size }& {\small Avg call time}   & {\small Avg Trn time}\\ \hline
    0.2    & 83 & 44 MB   & 115 KB &  \SI{6.0}{\micro\second} & 1 sec \\
    0.5  & 85 &42 MB   & 115 KB &  \SI{6.0}{\micro\second} & 1 sec \\ 
    0.9 & 57 &33 MB &  115 KB  &   \SI{5.6}{\micro\second}& 1 sec \\ 
    \hline
\end{tabular}
    \label{tab:training_dataset_size}
\end{table}

\noindent  \textbf{Total size.} 
Fig \ref{fig:diff-zipfian-50M} shows total size of all KV stores after all write requests are issued.
DumpKV achieves similar total size compared to RocksDB-std, TerakDB and Titan. This shows that DumpKV can accurately build short and long lifetime threshold based on workload and give relatively accurate prediction of lifetime of keys. 
Noted that RocksDB struggles to achieve low total size because it does not check validity of keys by searching LSM-tree during garbage collection. And its garbage definition does not reflect real number of invalidated keys.  

\noindent \textbf{Throughput.} Fig \ref{fig:diff-zipfian-50M} shows throuput of all KV stores.
Dumpkv achieves 25\%-40\% higher write rate copmared to RocksDB because it does not need to write large value during compaction which also includes garbage collection.
RocksDB-std shows highest write rate because it does not need to write KV pairs to SSTables and value files separately during flush process.   DumpKV achieves 5\%-13\% lower write rate performance compared to Titan and Terakdb because of code related implementation optimization.

\textbf{Feature storage overhead}
Fig \ref{fig:diff-zipfian-50M} shows LSM-tree size of all KV stores.
Compared to RocksDB, DumpKV has 14\%-18\% LSM-tree size increase. This LSM-tree size increase only accounts for 0.05\% total size increase which is small overhead because large values take up the majority of storage space.
Noted that LSM-tree size is small so that the whole LSM-tree can fit into block cache entirely. 

\subsection{Model overhead and feature importance}
We analyze model training and inference overhead  of DumpKV.
Fig \ref{fig:model_overhead} shows model calling time during garbage collection process for different workload. It  takes about 10\% of total garbage collection time to do  model calling to give remaining lifetime prediction to valid keys which is small overhead. DumpKV only invokes model to give remaining lifetime prediction if KV pairs in value files are still valid which leads to less computation overhead.   

Table \ref{tab:training_dataset_size} shows model training count, average training dataset size, model size, average model calling time and average training time for workload with different skewness.
Training count for workload that is less skewed is higher than that of more skewed workload. This is because less skewed workload tend to have more  
balanced labelled training datasets.
Average dataset size for each workload ranges from 33MB to 44MB which is small memory overhead. Workload with 0.2 skewness costs largest amount of memory because training samples have more deltas in average in less skewed workload.
The memory footprint of the GBM model, which is a mere 115KB across all workloads, is deemed a negligible overhead in the context of this study.
For training computation overhead,  it costs less than 1 second for each round of model training. 

Fig \ref{fig:feature_importance}(b) shows feature importance contribution to overall model training process. Feature $EDWC$ contributes 60\%-70\% to overall model performance gain which is highest among all features. This shows that write access frequency information is most useful in capture lifetime patterns of keys. The second highest contribution to model performance gain comes from $Deltas$ which takes up 16\%-25\%. This shows that past update distance information can be a useful feature for lifetime prediction as well.
Static features contribute 7\%-9\% of overall model performance gain.

  \begin{figure}[t]
    \begin{subfigure}[b]{0.22\textwidth}
        \resizebox{\linewidth}{!}{
            \includeinkscape[scale=0.44]{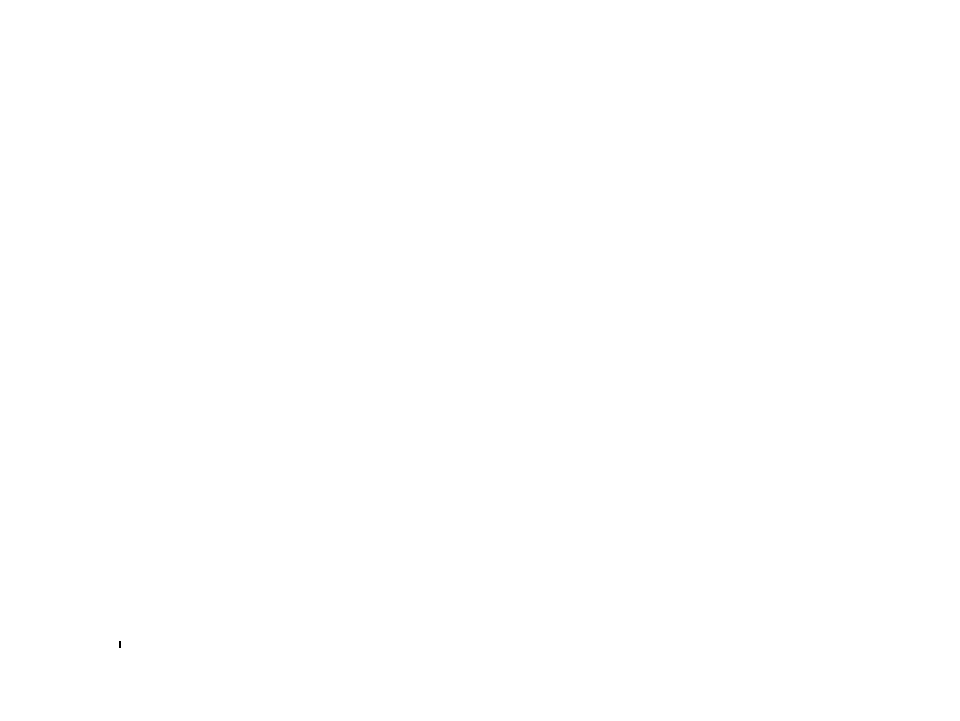}
        }
    \end{subfigure}
    \begin{subfigure}[b]{0.22\textwidth}
        \resizebox{\linewidth}{!}{
            \includeinkscape[scale=0.44]{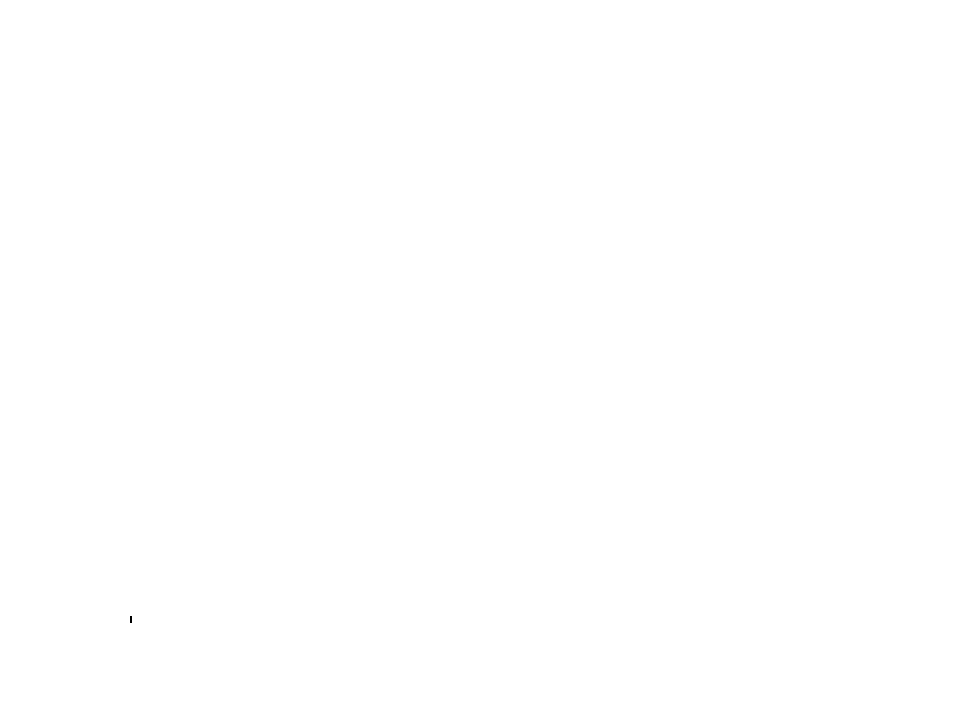}
        }
    \end{subfigure}
    \caption{Dynamic lifetime adjustment and GC rate change for 0.2 skewness   }
    \label{fig:lifetime_chg_0.2}
\end{figure}

  \begin{figure}[t]
    \begin{subfigure}[b]{0.22\textwidth}
        \resizebox{\linewidth}{!}{
            \includeinkscape[scale=0.44]{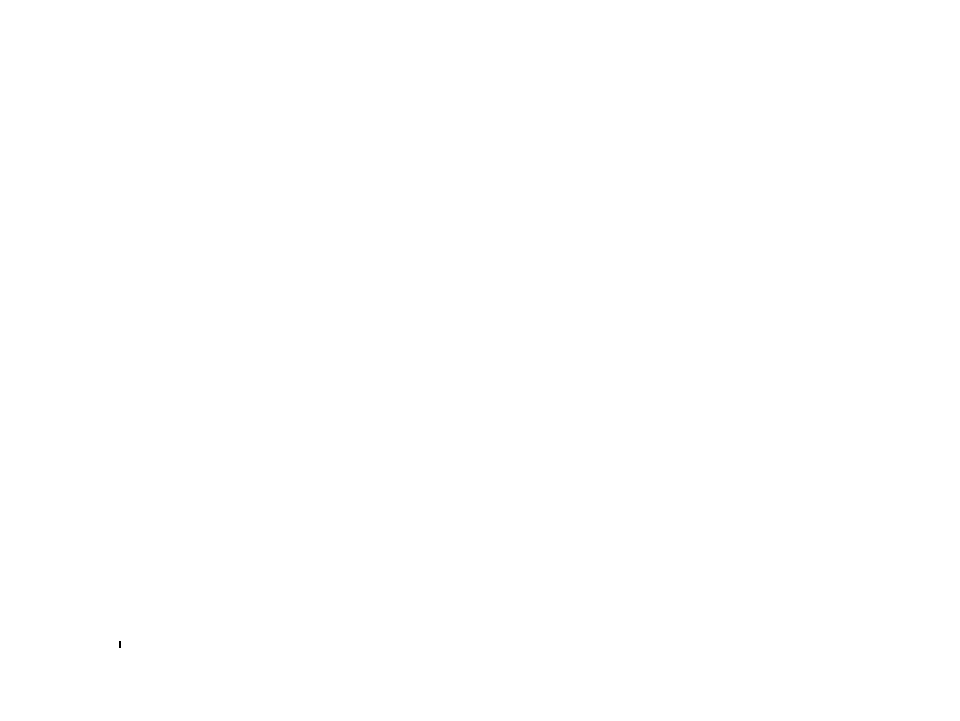}
        }
    \end{subfigure}
    \begin{subfigure}[b]{0.22\textwidth}
        \resizebox{\linewidth}{!}{
            \includeinkscape[scale=0.44]{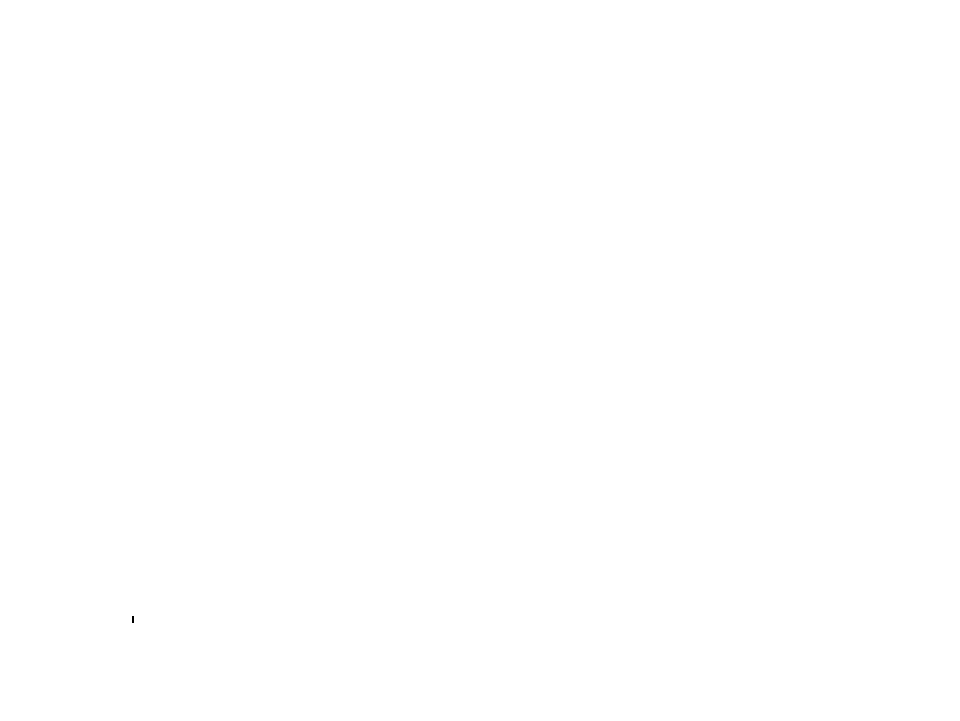}
        }
    \end{subfigure}
    \caption{Dynamic lifetime adjustment and GC rate change for 0.9 skewness }
    \label{fig:lifetime_chg_0.9}
\end{figure}

\subsection{Dynamic lifetime adjustment}
We next examine how dynamic lifetime adjustment in DumpKV helps determine short and long lifetime threshold based on garbage collection invalid ratio.
Fig \ref{fig:lifetime_chg_0.2} and Fig \ref{fig:lifetime_chg_0.9} 
shows  lifetime threshold for default, short and long lifetime value files  during workload running.   

Fig \ref{fig:lifetime_chg_0.2} and Fig \ref{fig:lifetime_chg_0.9}  shows garbage collection invalid ratio for default, short and long lifetime value files. Higher invalid ratio means more keys are invalid or more garbage are present during garbage collection.
DumpKV  dynamically adjusts default, short and long lifetime threshold for value files based on garbage collection invalid ratio. 
For workload with 0.9 zipfian skewness, DumpKV shows larger long lifetime threshold value compared to that in workload with 0.2 zipfian skewness. This shows that DumpKV is able to dynamically adjust lifetime threshold of value files  for different workloads with lifetime distribution monitoring. 
Note that default lifetime value files show highest garbage collection invalid ratio 0.72 and 0.75 which means majority of KV pairs are invalidated in default lifetime window. Default lifetime threshold and short lifetime threshold depicts similar trend because of high garbage collection invalid ratio for default lifetime value files.

  \begin{figure*}[t]
        \resizebox{\linewidth}{!}{
            \includeinkscape[scale=0.40]{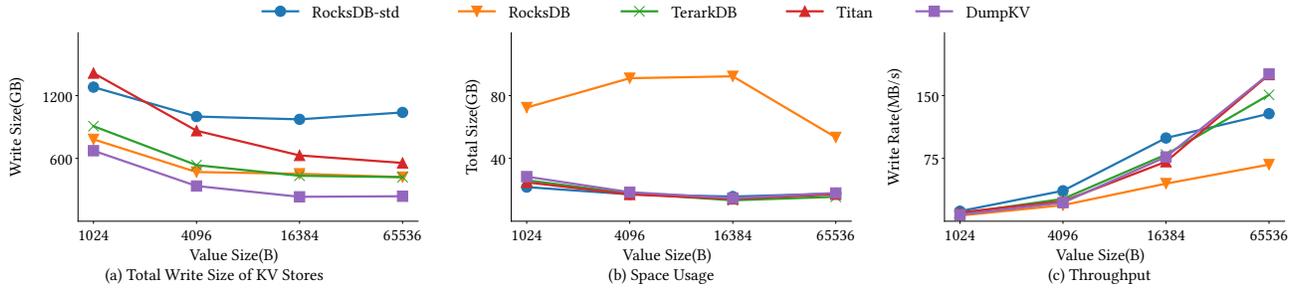}
        }
    \caption{Performance of KV stores under different KV pair sizes}
    \label{fig:varying_size}
  \end{figure*}

\subsection{Varying value size}
We evaluate performance of DumpKV for different value sizes. Specifically, we compare all KV stores with value sizes ranging from 1KB to 64KB.
Total size of KV pairs loaded into all KV stores is 200GB for all value sizes. 
Initial default lifetime for value files are set to 20\% of total writes to accumulate training data for model training for DumpKV. Workload data is generated with YCSB and zipfian constant is set to 0.99 for all value sizes.
Fig \ref{fig:varying_size} shows results of total write size and total storage size of all KV stores for varying sizes.
DumpKV reduces total write size by 52\%-77\%, 22\%-48\%, 33\%-46\%, 57\%-66\% compared to RocksDB-std, RocksDB, TerakDB and Titan, respectively.
Titan shows highest total write size for value size 1024 because it initiates garbage collection when garbage ratio reaches threshold and total write size even surpass that of RocksDB-std which means that picking garbage collection triggering ratio is challenging.
RocksDB-std struggles to achieve low total write size because it keeps writing large value to LSM-tree during compaction process for all value sizes.

When comparing total sizes, DumpKV exhibits a marginally larger size than both TerakDB and Titan, with an increase of approximately 10\%-15\% and 0.2\%-17\% , respectively. This mainly comes from LSM-tree storage size increase caused by extra feature storage overhead. RocksDB shows 59\%-84\% higher total size than DumpKV because it can not reclaim storage space of invalidated KV pairs in value files in time due to the fact that it does not check validity of KV pairs during garbage collection.  
When compared solely with the size of the LSM-tree, DumpKV exhibits a size increase ranging from 12\% to 17\% in comparison to RocksDB.

DumpKV shows similar throughput compared to Titan and TerakDB for all value sizes. DumpKV depicts 11\%-22\% throughput decrease compared to RocksDB-std for value sizes ranging from 1024 to 16384. 

Overall, DumpKV achieves lower write amplification with efficient garbage collection with its lifetime aware learning based garbage collection policy without sacrificing storage space and performance.

\subsection{With Model vs. Without Model}

  \begin{figure}
    \begin{subfigure}[b]{0.23\textwidth}
        \resizebox{\linewidth}{!}{
            \includeinkscape[scale=0.44]{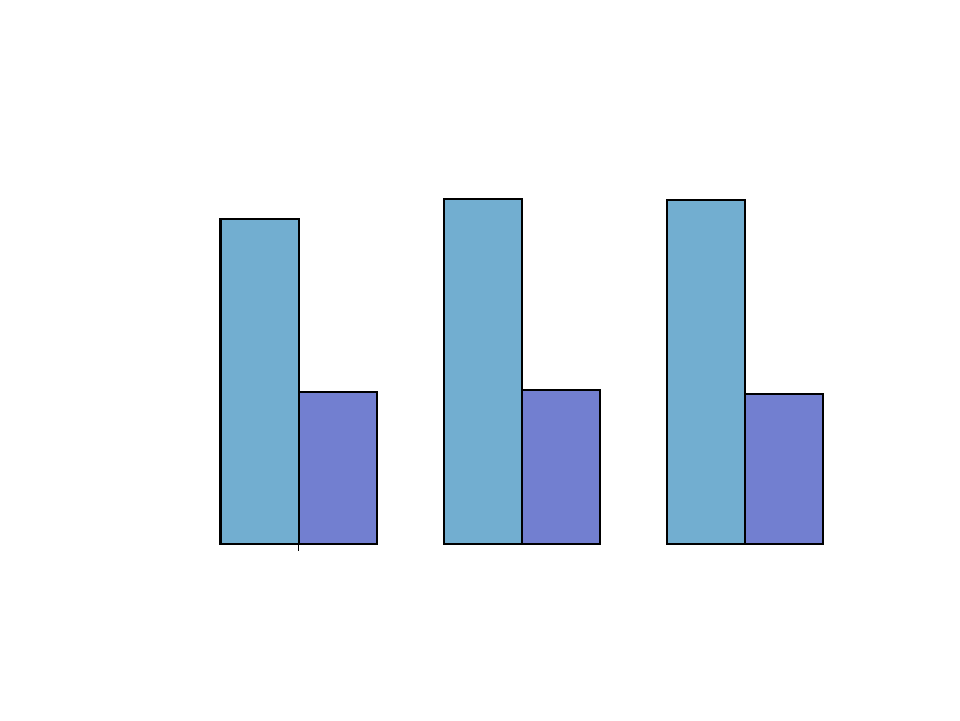}
        }
        \caption{(a) Total size   }
    \end{subfigure}
    \begin{subfigure}[b]{0.23\textwidth}
        \resizebox{\linewidth}{!}{
            \includeinkscape[scale=0.44]{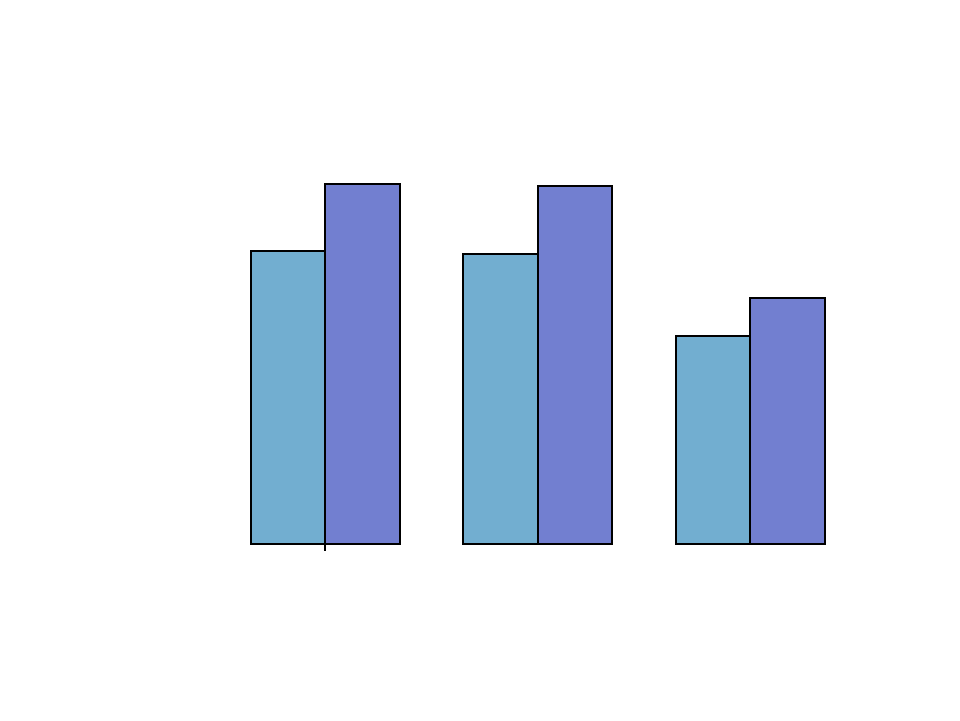}
        }
       \vspace{-11pt} 
        \caption{(b) Total write size }
    \end{subfigure}
    \caption{With model and without model for workload with different skewness}
\end{figure}

We analyze impact of introducing model to give remaining lifetime for KV pairs during garbage collection to show effectiveness of model prediction.

We compare total write size and total size of experiment results from enabling and disabling model calling. 
Specifically, when model is disabled all valid KV pairs during garbage collection are given long remaining lifetime prediction. Other parameters for model training and inference is the same as above.
Fig shows experiment results from different skewness workload ranging from 0.2 to 0.9. 
Model enabled DumpKV achieves 53\%-56\% lower total size compared to model disabled version. Total write size for model enabled DumpKV is 18\%-23\% higher than model disabled versino. This shows that DumpKV can relatively  give effectively accurate remaining lifetime prediction for KV pairs and model learns effective features from training dataset.

\section{\textbf{Discussion}}
In this work we use features that is derived from past write access information from individual keys. Features such as $Deltas$ and $EDWC$ are general features and can be applied to various application scenarios. 
Other features that are related to application specific scenarios can be integrated into DumpKV to boost prediction accuracy during garbage collection. 
For example, for e-commerce storage scenario user gender and order frequency can be leveraged as feature for each write.
For search log storage scenario, user location and user preference can be used as features for model training and inference.
Keys write access correlation can also be leveraged to capture lifetime correlation of write requests of keys, thus helping model to give more accurate lifetime prediction for keys during garbage collection. For example, keys in the same range might share similar write access patterns which means lifetime of keys in the same key range group is similar.

Other feature collection approach can be used to achieve more efficient feature engineering. For example, during garbage collection keys that are dropped and keys that are still valid can be associated with each other to discover write access and lifetime pattern for keys that are garbage collected in the same time to find out better feature value that can distinguish between each other more efficiently.

\section{\textbf{Related Work}}

\noindent \textbf{Garbage collection}
Lifetime information is used for garbage collection in fair amount work for storage hardware such as SSD and Hdd and file system \cite{midas, pcstream, sepbit}. 
MiDAS \cite{midas} introduces a systematic approach that reduces garbage collection overhead in log-structured systems by dynamically adjusting groups based on workload patterns, leading to lower WAF, higher throughput, and less memory and CPU usage.
PCStream \cite{pcstream} proposes a fully automatic stream management technique for multi-streamed SSDs that improves IOPS  and reduces garbage collection overhead by  identifying dominant I/O activities and effectively doubling the number of available streams using a new type of streams called internal streams.
This idea is also found in AutoStream\cite{autostream} and FStream \cite{fstream}
SepBIT \cite{sepbit} introduces a novel data placement algorithm, reduces write amplification in log-structured storage by inferring block invalidation times from workloads and grouping blocks with similar estimated times, thereby improving I/O throughput. 
Several research works also study how to reduce or avoid garbage collection overhead \cite{iplfs, real-time-gc, heuristic-clean, io-gc}.
IPLFS \cite{iplfs} develops a garbage collection-free log-structured filesystem with Interval Mapping for efficient LBA-to-PBA translation.
A fair amount of work proposes to perform garbage collection in idle period or in an preemptive way\cite{improve-log-structure-adaptive, suspend-aware, semi-premp}.

\noindent \textbf{LSM-tree based KV stores}
Multiple studies have been done to improve read write performance of LSM-tree based storage engine \cite{silk, triad, lsm-tree-local-cloud}.
NoveLSM \cite{novelsm} uses non-volatile memories and techniques like byte-addressable skip list, direct mutability, and read parallelism to reduce write and read latency. 
Enabling times and persistent deletion in lsm-tree engine
Calcspar \cite{calcspar} proposes a contract-aware LSM store for cloud storage, which addresses latency variances of Amazon EBS by regulating I/O requests and uses a fluctuation-aware cache and a congestion-aware IOPS allocator to tail latency and average latency.
DiffKV \cite{diffkv}  uses KV separation and fine-grained KV differentiation by size to manage keys and values to achieve high performance in writes, reads, and scans.
ADOC \cite{adoc} proposes a tuning framework that minimizes data overflow and reduces write stalls by automatically adjusting system configurations to reduce write stall.
A fair amount of work is done to reuce write amplification of LSM-tree \cite{vttree, pebblesdb, dosto}.
MatrixKV \cite{matrixkv} proposes multi-tier DRAM-NVM-SSD systems, reduces write stalls and amplification by performing smaller L0–L1 compactions and reducing LSM-tree depth. 
Wisckey \cite{wisckey} introduces 
HashKV \cite{hashkv} uses hash-based data grouping for efficient updates and garbage collection under update-intensive workloads, achieving higher throughput and  less write traffic.

\noindent \textbf{Use of machine learning to improve system performance }
A fair amount of work has also been done by leveraging machine learning model in storage system \cite{chess, lrb, leaper}. 
LinnOS \cite{linnos} uses a light neural network to infer SSD performance per-IO to improve average I/O latencies accuracy.
GL-Cache \cite{glcache}  clusters similar objects and performs learning and eviction at the group level, achieving higher efficiency and throughput.
Leaper \cite{leaper} predicts and prefetches hot records in LSM-tree storage engines, addressing cache invalidation issues caused by background operations  to reduce cache invalidations and latency spikes .
Machine learning is also adopted in query optimizer \cite{microlearner} and database tuning \cite{auto-tune, qtune}.
Learning approach is also adopted in learned index in storage system \cite{alex-learned-idx, kipf2019sosd, sagdb}.
XSTORE \cite{xstore} uses a machine learning model as a learned cache for the tree-based index to improve performance and scalability and outperforms state-of-the-art RDMA-based stores and reduces client-side memory usage.
Several recent works also use machine learning to do cache prefetching \cite{baleen, cminer, context-prefetch, mithril}.
 Baleen \cite{baleen} uses machine learning and a new cache residency model called episodes, to reduce peak backend load and total cost of ownership  by optimizing the Disk-head Time metric and managing flash write rate.

\section{\textbf{Conclusion}}
In this paper we introduce DumpKV, a learning based lifetime aware garbage collection framework for KV separation in LSM-tree. 
DumpKV proposes to collect training samples from $L0-L1$ compaction and garbage collection process by leveraging the fact that LSM-tree is relatively small so that majority of LSM-tree can be stored in block-cache which makes features generation is fast. 
DumpKV  predicts remaining lifetime of keys during garbage collection to balance between throughput, write amplification and total size. 
DumpKV also adjust lifetime threshold for value files dynamically to adapt to workload shift.  
DumpKV is the LSM-tree storage engine system that applies learning based approach to leverage lifetime information to improve garbage collection efficient for KV separation architecture. 
Evaluation results show that DumpKV reduces total write size by 38\%-73\% compared to other KV stores acorss different skewed workload and achieves comparable results in terms of total size and throughput.

\bibliographystyle{ACM-Reference-Format}
\bibliography{sections/dumpkv}

\appendix

\end{document}